\documentstyle[12pt]{article}
\renewcommand
\baselinestretch{1.4}

\def\beq{\begin{equation}}
\def\brr{\begin{array}}
\def\err{\end{array}}
\def\eeq{\end{equation}}
\def\bea{\begin{eqnarray}}
\def\eea{\end{eqnarray}}
\def\bs{\bigskip}

\def\nn{\nonumber}

\begin{document}

\vspace*{25mm}

\begin{center}

{\large \bf THE VACUUM ENERGY DENSITY FOR SPHERICAL AND
CYLINDRICAL
UNIVERSES}

%or SEARCH FOR THE MOST STABLE QUANTUM NO-BOUNDARY UNIVERSE

\vspace{4mm}

\renewcommand
\baselinestretch{0.8}
\medskip

{\sc E. Elizalde}\footnote{On leave of absence from and
permanent address: Department E.C.M., Faculty of Physics,
Barcelona University, Diagonal 647, E-08028 Barcelona, Spain;
e-mail: eli @ ebubecm1.bitnet}
\\ {\it Institute of Theoretical Physics, Chalmers University of
Technology, \\
 S-412 96 G\"oteborg, Sweden}

\renewcommand
\baselinestretch{1.4}

\vspace{5mm}

{\bf Abstract}

\end{center}

The vacuum energy density (Casimir energy) corresponding to a
massless scalar quantum field living in different
universes (mainly no-boundary ones), in several dimensions, is
calculated.
Hawking's zeta function regularization procedure supplemented with
a very simple binomial expansion is shown to be a rigorous and well
suited method for performing the analysis. It is compared with other,
much more involved techniques. The principal-part prescription
is used to deal with the poles that eventually appear.
Results of the analysis are the absence of poles at four dimensions (for
a 4d Riemann sphere and for a 4d cylinder of 3d Riemann spherical
section), the total
coincidence of the results corresponding to a 3d and a 4d cylinder
(the first after pole subtraction), and the fact that the
vacuum energy density for cylinders is (in absolute value) over an order
of magnitude smaller than for spheres of the same dimension.

\vspace{3cm}

%\noindent PACS numbers: 03.70.+k, 04.20.Cv, 11.10.Gh

\newpage

\section{Introduction}

The investigation of the vacuum energy density (or Casimir energy)
\cite{cas} corresponding to a quantum field theory defined in a certain
manifold (spacetime) with or without boundary, is one of the most
basic issues of quantum field theory. The Casimir force is probably
the simplest and most spectacular of the different manifestations
of the vacuum energy ---better, of the modification of the vacuum
energy when boundaries, curvature or a background field are
superimposed to the manifold \cite{cas,bvw}. To understand the
correspondence between the {\it sign} of the effect (associated with an
attractive or repulsive force, depending on the sign being negative
or positive, respectively) and the specific topology of the
boundary is a most challenging point which requires explanation.
A historical failure in relation with this issue ocurred some years
ago, when the Casimir force was thought to contain the explanation
of quark confinement in the context of the bag model. The idea was that
this vacuum effect would provide the attractive force
giving tension to the bag that contained the quarks. But it turns out
that the calculation of the force for curved boundaries is very
tricky, plagued with infinities and needing apropriate regularization.
Sometimes cut-offs remain, zeta-function regularization is
said to be insufficient, and it is difficult to extract uncontroverted,
physically meaningful results \cite{bvw}. Actually, this subject has
been
the source of several sound errors in the scientific literature. In any
case, when it was proven, without doubt, that the force for a
closed sphere in three-dimensional space is {\it repulsive} (as it
is for a closed cube, and not attractive as for a
pair of plates) the brilliant idea of supplementing the bag model
with the Casimir force had to be abandoned.

Notwistanding that, the presence of the Casimir force in very
different phenomena of condense matter, solid state and laser
physics has been rigorously stablished, both theoretically and
experimentally \cite{cole}. Its relevance for possible models of our
universe is also without discussion. Maybe owing to the difficulties
encountered when trying to give a plausible answer to the question:
{\it what are the boundary conditions of our universe?}, some of
the most popular models of spacetime nowadays are given by
manifolds without boundaries.

Riemann spheres are to be counted among the simplest and most
important of these manifolds.
This is the reason why we have chosen them as the spaces  of
our models. We will study these models in different dimensions, with
the hope to find some characteristic that may singularize some of
the manifolds considered and some particular dimension, among all
of them.

{}From a more technical point of view, the specifications of our
study are as follows. We shall use the most elegant, rigorous and simple
of the regularizations known to date, namely zeta function
regularization \cite{haw}. Moreover, we are going to supplement it with
a most easy  technique, which is binomial expansion \cite{act1}. Then we
shall compare
our findings with those coming from other approaches, which turn out to
be much more artificial, lengthy and less rigorous. Hard to believe
as this statement may seem, it is the plain truth. Little merit
about it, again this is just a confirmation of the famous
general principle, atributed to Einstein, which asserts that nature
always follows the most simple path among the ones that are available to
her.
That such path turns out to be in this case, at the same time, the
most rigorous mathematically is however a rewarding small surprise
(since sometimes physical intuition and beauty has been associated
with mathematical sloppiness). Even from the pure calculational point
of view,  the method here developed yields  very
rapidly convergent expressions which allow us to obtain 6 or 8 digit
precission with just a few first terms of a series, in a home computer
and with a standard computation package ({\it Mathematica}, for
instance, but {\it never} use it for dealing with the Hurwitz zeta
function $\zeta (s,a)$ when $a>1$). As side products of our
analysis, some asymptotic expansions for Hurwitz and Epstein zeta
functions, that had been obtained by the author previously, are here
chequed and numerically contrasted with other results in the
literature. As has been pointed out already, this is a necessary
exercise in this field, because of the discrepancies and errors
that
so frequently appear. In particular, a small table of derivatives
of the Riemann zeta function, which are repeatedly used in our
numerical calculations, is given.

In Sect. 2 we summarize our method of zeta function regularization
and describe the way in which the binomial expansion is used. We
compare it with other approaches that have been employed, in other
to prove the advantages of our procedure. In Sect. 3 we obtain the
Casimir energy density for Riemann spheres in $d=1,2,3,4$
dimensions. These are manifolds without boundary, but also the situation
when one has a half such manifold with Dirichlet and Neumann boundary
conditions, respectively, is  considered for comparison.
Cylindrical manifolds whose sections are Riemann spheres, with
total dimension $d=2,3,4,5$ are investigated in Sect. 4. The
discussion of our results and the conclusions of our analysis are
given in Sect. 5.
\bs

\section{Analytical approach to the zeta functions for Riemann
surfaces}

As has been observed elsewhere \cite{eku1}, the formula
\cite{ejmp} (see also \cite{brecl})
\bea
F(s;a,b) & \equiv & \sum_{n=1}^\infty \left[ (n+a)^2 +b \right]^{-s} =
\frac{b^{-s}}{\Gamma (s)} \sum_{n=0}^\infty \frac{(-1)^n \Gamma
(n+s)}{n!} \, b^{-n} \zeta (-2n, a)  \\
&& + \frac{\sqrt{\pi}\, \Gamma (s - 1/2)}{2 \Gamma (s)} \, b^{1/2 -s}
+ \frac{2 \pi^s}{\Gamma (s)} \cos (2 \pi a) \, b^{1/4 -s/2}
\sum_{n=1}^\infty n^{1/2 -s} K_{s-1/2} (2 \pi n \sqrt{b} ), \nn
\label{eh1}
\eea
and its generalizations to higher dimensional manifolds
\cite{ejmp}, are extremely useful (owing to exponentially quick
convergence) when $b >0$. However, it turns out to be rather
difficult to apply them when $b<0$.
And this is usually the relevant case which appears when one tries
to calculate, e.g., the determinants of the Laplacian operators on
Riemann spheres by using zeta functions. One has to regularize
(i.e., analytically continue) expressions of the general form
\beq
f(s; a,b,c) \equiv \sum_{l=1}^\infty l^{-s+b} (l+a)^{-s+c},
\label{gfo}
\eeq
where $a$ turns out to be positive.

An alternative expression to (\ref{eh1}) has been obtained in
\cite{wcmp}, by making use, essentially, of the usual integral
representation for the Hurwitz zeta function together with the
Mellin transformation (actually the same ingredients, aside from
adequate series commutation, which were used in \cite{ejmp} for the
derivation of (\ref{eh1})). Specifically, for
\beq
\zeta^{(n)} (s) \equiv \sum_{l=1}^\infty \left[ l^{-s} (l+2n+1)^{1-
s} +l^{1-s} (l+2n+1)^{-s} \right],
\eeq
the following interesting result was obtained
\beq
\left. \frac{d}{ds} \zeta^{(n)} (s) \right|_{s=0} = 4 \zeta'(-1) -
\frac{1}{2} (2n+1)^2 + \sum_{k=1}^{2n+1} (2k-2n-1) \log k,
\eeq
from which, in particular, for the zeta function of the Laplacian
on the hemisphere with Dirichlet and Neumann boundary conditions,
respectively,
\beq
\zeta_D (s) = \sum_{l=1}^\infty l [l(l+1)]^{-s}, \ \ \ \ \  \zeta_N
(s) = \sum_{l=1}^\infty (l+1) [l(l+1)]^{-s},
\label{zdn}
\eeq
one gets
\beq
\zeta_D' (0) =2 \zeta' (-1) + \frac{1}{2} \log (2 \pi) -
\frac{1}{4}, \ \ \ \
\zeta_N' (0) = 2\zeta' (-1) - \frac{1}{2} \log (2 \pi) -
\frac{1}{4}.
\label{dn2}
\eeq
Those are nice results, indeed. However, the derivation of these
expressions in \cite{wcmp} is not free from difficulties. In fact,
to start with, the analysis is rather lengthy and, on the other
hand, a highly arbitrary, additional regularization is needed at
some point: a certain infrared convergence factor $t^s$ ($t$ is the
integration variable) must be introduced and the exponent $s$ can
be let to go to zero only after performing a convenient combination
of different terms. That these manipulations are not so obviously
accepted (even if in the end they turn out to be right, as is the
case here) is proven by the continuous recourse to specific checking of
the final numbers with well-known results \cite{wcmp,psa,itz}. In
certain cases, in fact, use of these integral transforms can lead to
discrepancies with already known results. So, the value of
$\zeta'(-1)$ obtained in \cite{nas} as a byproduct of an original
method there developed seems to be quite far from the best accepted
value
as given, for instance, in \cite{elfr} (see expressions (\ref{zpr})
below). Our value is coincident (at least to 8 digits) with
the one obtained by Salomonson using a different formula
\cite{psal2}. On the other hand, however, the remarkable expression
obtained in \cite{nas} for $\zeta'(-2)$ appears to be right (see
(\ref{zpr})).

A mathematically clean, rigorous and, at the same time, much more
simple procedure to deal with any expression of the general form
(\ref{gfo}) can be devised which makes use of the simplest of
ideas: binomial expansion \cite{act1}. This goes as follows
\bea
f(s; a,b,c)& \equiv &\sum_{l=1}^\infty l^{-s+b+c} (l+a)^{-s+c} =
\sum_{l=1}^\infty l^{-2s+b} (1+al^{-1})^{-s+c}  =
\sum_{l=1}^{[a]} \{ \ \ \} + \sum_{l=[a]+1}^\infty \{ \ \ \} \nn \\ &=&
g(s; a,b,c) +  \sum_{l=[a]+1}^\infty \sum_{k=0}^\infty \frac{\Gamma
(1-s+c)}{k! \Gamma ( 1-s-k+c)} a^k l^{-2s-k+b+c},
\eea
being $[a]$ the integer part of $a$, so that g(s) is an integer
function of $s$, while the second, truncated series is absolutely
convergent (since $al^{-1} <1$ there). The final result is
\bea
f(s; a,b,c) = \sum_{l=1}^{[a]} l^{-2s+b+c} (1+al^{-1})^{-s+c}
+\sum_{k=0}^\infty \frac{\Gamma (1-s+c)}{k! \Gamma ( 1-s-k+c)} a^k
&& \nn \\ \times  \left[ \zeta (2s+k -b-c) -  \sum_{l=1}^{[a]}
l^{-2s-k+b+c} \right]. &&
 \eea
In particular,
\beq
f(0; a,b,c) = \sum_{l=1}^{[a]} l^b (1+al^{-1})^c +\sum_{k=0}^\infty
\frac{\Gamma (1+c)}{k! \Gamma ( 1-k+c)} a^k
\left[ \zeta (k -b-c) -
\sum_{l=1}^{[a]} l^{-k+b+c} \right],
\eeq
and
\bea
 f'(0; a,b,c) &=&- \sum_{l=1}^{[a]} \left[ 2\log l+ \log (1+al^{-1})
\right] l^b (1+al^{-1})^c \nn \\ &+&\sum_{k=0}^\infty \left\{ \frac{\psi
(1- k+c) - \psi (b)}{\Gamma ( 1-k+c)} \,
 \left[ \zeta (k -b-c) -  \sum_{l=1}^{[a]} l^{-k+b+c} \right] \right.
\\ &+& \left. \frac{2\Gamma (1+c)}{\Gamma (1-k+c)}
\left[ \zeta' (k -b-c) +
\sum_{l=1}^{[a]} l^{-k+b+c} \log l \right] \right\}
\frac{a^k}{k!}, \nn
\eea
In fact this last formula is a bit tricky, and has to be modified
(in general) when
$b$ and $c$ are integers. Then, the derivative for the particular
value $k=b+1$ must be done with special care, by performing first
expansions around the poles and zeros of these functions at $s=0$
(this will be shown in full detail below).

For the zeta function $\zeta_D (s)$, eq. (\ref{zdn}), we obtain
\beq
\zeta_D (s) = \sum_{l=1}^\infty l^{1-2s} (1+l^{-1})^{-s}=2^{-s} +
\sum_{k=0}^\infty \frac{\Gamma (1-s)}{k! \Gamma ( 1-s-k)} \left[
\zeta (2s+k-1) -1 \right],
\eeq
and
\beq
\zeta_D' (0) = 2\zeta' (-1) +\frac{5}{4} +\frac{\gamma}{2} -\log 2
-\sum_{k=2}^\infty (-1)^k \ \frac{\zeta (k) -1}{k+1}.
\label{mid}
\eeq
The power of our simple method will be now demonstrated by showing
that this last expression coincides with the first of (\ref{dn2}).
In fact, we have
\beq
\sum_{k=2}^\infty (-1)^{k+1} \frac{\zeta (k)}{k+1}
=\sum_{k=2}^\infty \frac{(-1)^{k+1}}{(k+1)\Gamma (k)}
\int_0^\infty dt \,  \frac{t^{k-1}}{e^t -1} \equiv \varphi (1),
\eeq
with
\beq
\varphi (u) \equiv \int_0^\infty \frac{dt}{e^t-1} \sum_{k=2}^\infty
\frac{(-u)^{k+1}t^{k-1}}{(k+1) \ (k-1)!}.
\eeq
It is easy to see that $\varphi'(u)= u [\psi (1) -\psi (u+1)]$ and
integrating, $\varphi (1)=- \gamma /2 -1 + (1/2) \log (2 \pi)$.
Finally, adding the rest of the series:
\beq
\sum_{k=2}^\infty \frac{(-1)^{k}}{k+1} =- \frac{1}{2} + \log 2,
\eeq
we obtain the desired result, i.e. that (\ref{mid}) coincides with  the
first of eqs. (\ref{dn2}). The numerical value is
\beq
\zeta_D' (0) = 0.338096.
\eeq

A second particular example is the following. For the case of a
rectangle (of sides $a$ and $b$) with Dirichlet boundary
conditions, the spectrum of the Laplacian is $\lambda_{mn} = \pi^2
(m^2/a^2 + n^2/b^2)$, and the zeta function
\bea
\zeta_{rec} (s) &=& \pi^{-2s} \sum_{m,n=1}^\infty \left(
\frac{m^2}{a^2} + \frac{n^2}{b^2} \right)^{-s} = -\frac{1}{2}
\left( \frac{b}{\pi} \right)^{2s} \zeta (2s) +\frac{a}{2\sqrt{\pi}}
\left( \frac{b}{\pi} \right)^{2s-1} \frac{\Gamma (s-1/2)}{\Gamma
(s)} \, \zeta (2s-1) \nn \\
&&+ \frac{2}{\Gamma (s)} \left( \frac{ab}{\pi} \right)^s
\sqrt{\frac{a}{b}} \sum_{m,n=1}^\infty \left( \frac{m}{n}
\right)^{s-1/2} K_{s-1/2} (2\pi mn a /b),
\eea
where, again, the corresponding asymptotic expansion for the
Epstein zeta function in \cite{ejmp} has been used.
Taking now the derivative, we get
\beq
\zeta_{rec}' (0) = \frac{1}{2} \log (2b) + \frac{\pi a}{12  b} +2
\sqrt{\frac{a}{b}} \sum_{m,n=1}^\infty \sqrt{ \frac{n}{m}} \ K_{-
1/2} (2\pi mn a /b),
\label{z0r}
\eeq
which is best for numerical computations when $a\geq b$. In the
particular case $a=b$ (square), this reduces to
\beq
\zeta_{sq}' (0) = \frac{1}{2} \log (2a) + \frac{\pi }{12} +2
\sum_{m,n=1}^\infty \sqrt{ \frac{n}{m}} \ K_{-1/2} (2\pi mn),
\label{z0s}
\eeq
which is just another expression for the same result obtained in
\cite{psa} (cf. eq. (A11.3))
\beq
\zeta_{sq}' (0) = \frac{1}{2} \log (2a) +\frac{1}{4} \log (8\pi)
+\frac{1}{2} \log  \frac{\Gamma (3/4)}{\Gamma (1/4)}.
\eeq
The numerical value is, in both cases,
\beq
\zeta_{sq}' (0) = \frac{1}{2} \log (2a) + 0.263672.
\eeq
Notice, however, that for the general rectangle, expression (\ref{z0r})
is of much more practical use than the well-known one in terms of
Dedekind's modular form $\eta$, i.e. \cite{psa,itz}
\beq
\zeta_{rec}' (0) = \frac{1}{4} \log (ab) -\log \left[
\frac{1}{\sqrt{2}} \left( \frac{b}{a} \right)^{1/4} \eta (q)
\right], \ \ \ \ \eta (q) = q^{1/24} \prod_{m=1}^\infty (1-q^m), \
\ \ \ q=\exp \left( -2\pi \sqrt{\frac{b}{a}} \right).
\eeq
In fact, a  few first terms of the series in  (\ref{z0r}) suffice
to obtain extremely accurate numerical results (just as for the case of
the square).
 \bs

\section{The vacuum energy density for Riemann surfaces in
different dimensions}

We shall here calculate the vacuum energy density corresponding to
a massless scalar field  living in a Riemann sphere
(manifold without boundary) or in a part thereof (namely a half-sphere,
with Dirichlet or Neumann boundary conditions). The calculations
become rather simple and precise numerical results are easy to
obtain, by making exhaustive use of the formulas and considerations
of the preceding section. But rather than attributing it to the specific
manipulations we have carried out there (nothing special, in fact), this
simplicity
is to be interpreted as a success of the zeta-function regularization
procedure itself \cite{haw}.

\subsection{Dimension d=1}

 This case is rather trivial and deserves no comment. For the
one-dimensional Riemann sphere (no boundary, hence no boundary
conditions), the vacuum energy density (or Casimir energy) is
\beq
E_1 = \frac{\hbar}{2r^2} \, \sum_{n=1}^\infty 2 \sqrt{n^2}=
\frac{\hbar}{r^2} \, \sum_{n=1}^\infty n = \frac{\hbar}{r^2}
\,\zeta (-1),
\eeq
being $r$ (here and in what follows) the radius of the Riemann
sphere, and
\beq
  \zeta (s) =\sum_{n=1}^\infty n^{-s},
\eeq
just the ordinary Riemann zeta function;  therefore
 \beq
E_1 =- \frac{\hbar}{12r^2}.
\eeq

For the semicircumference, the eigenmodes are shared by the cases
of Dirichlet and Neumann boundary conditions, respectively. We obtain
\beq
E_1^D = E_1^N =- \frac{\hbar}{24r^2}.
\eeq

\subsection{Dimension d=2}
For the ordinary Riemann sphere (i.e., the one which appears in
string theory, see for instance \cite{psa,dhp}), we have, in the
case of the half-sphere with Dirichlet and Neumann boundary
conditions,
\beq
E_2^D = - \frac{\hbar}{2r^2} \sum_{i}^D m_i^D \lambda_i, \ \ \ \ \
E_2^N = - \frac{\hbar}{2r^2} \sum_{i}^N m_i^N \lambda_i,
\eeq
respectively, where the the eigenvalues $\lambda_i$ and eigenmode
multiplicities $m_i^D, \ m_i^N$ are
\beq
\lambda_i = \frac{\sqrt{l (l+1)}}{r}, \ \ \ \ m_i^D = l,  \ \ \ \
m_i^N = l+1.
\eeq
We get
\beq
E_2^{D,N} =\frac{\hbar}{2r^3} \zeta_2^{D,N} (-1/2), \ \ \  \
\zeta_2^D (s) = \sum_{l=1}^\infty l [l(l+1)]^{-s},  \ \ \  \
\zeta_2^N (s) = \sum_{l=1}^\infty (l+1) [l(l+1)]^{-s},
\eeq
and by using our method as described in the preceding section,
\bea
\zeta_2^D (s) &=&2^{-s}+ \sum_{k=0}^\infty \frac{\Gamma (1-s)}{k!
\Gamma (1-s-k)} [\zeta (2s +k -1) -1],  \nn \\  \zeta_2^N (s)
&=&2^{1-s}+ \sum_{k=0}^\infty \frac{\Gamma (2-s)}{k! \Gamma (2-s-k)}
[\zeta (2s +k -1) -1].
\eea
Now, when evaluating $\zeta_2^{D,N} (s=-1/2)$, a pole for $k=3$
appears, in each case. The usual prescription in zeta function
regularization is to take the principal part of the pole
\cite{bvw,kk,zboo}. With this in mind, it is easy to obtain
\beq
\zeta_2^D (-1/2) =  \frac{1}{32(s+1 /2)} + 0.033532, \ \ \ \ \ \
\zeta_2^N (-1/2) =  -\frac{1}{32(s+1/2)} - 0.298630.
\eeq
This yields for the energy
\beq
E_2^D = 0.016766 \cdot \frac{\hbar}{r^3}, \ \ \ \ \ \ \ E_2^N = -
0.149314 \cdot \frac{\hbar}{r^3}.
\eeq

For the whole, ordinary Riemann sphere (no boundary), we have
\beq
\zeta_2 (s) = \sum_{l=1}^\infty (2l+1) [l(l+1)]^{-s} = \zeta_2^D
(s) +\zeta_2^N (s) = -0.265096,
\eeq
and  thus we obtain
\beq
E_2 = -0.132548 \cdot \frac{\hbar}{r^3}.
\eeq

Summing up, we see that in $2+1$ spacetime dimensions, where space
is the ordinary Riemann sphere (no boundaries), the Casimir energy
density is {\it negative}. It is the sum of the energies
corresponding to two half-spheres, one with Dirichlet and the other
with Neumann boundary conditions on the one-dimensional boundary.
The poles have opposite sign, and they anihilate when performing the
sum. This is certainly an interesting result. It should be compared with
the Casimir energy in $3+1$ spacetime and boundary  conditions
imposed  on a two-dimensional spherical surface. It is clear that
the sign of the Casimir energy density for Riemannian manifolds can
have important global cosmological consequences in plausible models
of our universe. In fact, it could account for a lessening of the
expansion ratio of the universe.

\subsection{Dimension d=3}

The three-dimensional Riemann sphere is a manifold without
boundary that could perfectly well correspond to the spatial part of our
universe, as a whole. The eigenvalues of the Laplacian operator
are $\lambda_i^2=l(l+2)/r^4$, with degeneracies $m_i = (l+1)^2$.
Thus, the vacuum energy density for a massless scalar field is
given by
\beq
E_3 = - \frac{\hbar}{2r^4} \, \zeta_3(s=-1/2), \ \ \ \ \zeta_3(s)
= \sum_{l=1}^\infty (l+1)^2 [l(l+2)]^{-s}.
\eeq
We can write,
\beq
 \zeta_3(s) = \sum_{l=2}^\infty l^{2(1-s)} (1-l^{-2})^{-s} =
 \sum_{k=0}^\infty \frac{(-1)^k \Gamma (1-s)}{k! \Gamma (1-s-k)}
[\zeta (2s +2k -2) -1],
\eeq
and
\beq
 \zeta_3(-1/2) = \sum_{k=0}^\infty \frac{(-1)^k \Gamma (3/2)}{k!
\Gamma (3/2-k)}  [\zeta (2k -3) -1] = - \frac{1}{16(s+1/2)}-0.411502,
\eeq
which has a pole, for $k=2$. Doing as above, we obtain the
following numerical result:
\beq
E_3 = -0.205751 \cdot \frac{\hbar}{r^4}.
\eeq

\subsection{Dimension d=4}

For the four-dimensional Riemann sphere, the corresponding
eigenvalues and multipli\-ci\-ties are $\lambda_i^2=l(l+3)/r^6$ and
$m_i = (l+1)(l+2)(2l+3)/6$. The vacuum energy density is now
\beq
E_4 = - \frac{\hbar}{2r^5} \, \zeta_4 (s=-1/2), \ \ \ \ \zeta_4(s)
=\frac{1}{6} \sum_{l=1}^\infty (l+1)(l+2)(2l+3) [l(l+3)]^{-s}.
\eeq
We can write,
\bea
 \zeta_4(s) &=&\frac{1}{3} \sum_{l=1}^\infty u (u^2-1/4) (u^2 -
9/4)^{-s} = \frac{1}{3}
 \sum_{k=0}^\infty \frac{(-1)^k \Gamma (1-s)}{k! \Gamma (1-s-k)} \nn \\
&& \times \left( \frac{9}{4} \right)^k
[\zeta (2s +2k -3, 5/2)- \zeta (2s +2k -1, 5/2) /4 ],
\eea
being $u=l+3/2$ and $\zeta (s,a)$ Hurwitz's zeta function
\beq
\zeta (s,a) =  \sum_{n=0}^\infty (n+a)^{-s}.
\eeq
Again, nothing else has been done here but to apply the procedure
as described in the preceding section. We thus obtain
\beq
 \zeta_4(-1/2) = \frac{1}{3}
 \sum_{k=0}^\infty \frac{(-1)^k \Gamma (3/2)}{k! \Gamma (3/2-k)}
\left( \frac{9}{4} \right)^k
[\zeta (2k -4, 5/2)- \zeta (2k -2, 5/2) /4 ]=-0.424550.
\label{rs4}
\eeq
It might seem that the term $(9/4)^k$ could spoil convergence. This
is not true: as always, we get a quickly convergent series. This is
guaranteed in advance by the procedure itself, but it is rewarding
to check this property numerically and see explicitly the rapid
convergence (warning to the reader: one {\it cannot} use {\it
Mathematica} to compute this expression since the Hurwitz zeta
functions $\zeta (s,a)$ are very ill defined in this program for
values of $a>1$; this applies at least to version 2.0).
Another important surprise is the fact that expression (\ref{rs4})
is {\it finite}, no pole appears in this case (contrary to previous
situations, in (\ref{rs4}) there is no  principal part
reduction).

Finally,
\beq
E_4 = -0.212275 \cdot \frac{\hbar}{r^5}.
\eeq

Before closing this section, the following observation is in order.
As explained before, an alternative treatment of the  zeta
functions  above would be simply to split the polynomial in powers
of the summation indices and then use the method of \cite{wcmp}. It
is easy to check that this procedure is much more lengthy than the one
developed here. On the other hand, the cancellation of poles in
this method must be done explicitly (resorting to expansions around
all poles and zeros), while it is
immediate in our procedure (actually, no pole is ever formed). We
conclude
that the most direct way (the one we use) turns out to be here, at the
same time, the shortest, most rigorous and best suited for numerical
evaluation.

The numerical results corresponding to the different cases here
considered are depicted in Fig. 1. The vacuum energy density for a
massless scalar field living in Riemann spheres (no-boundary
manifolds) is represented as
a function of the space dimension, $d$, in units of $\hbar \
r^{-(d+1)}$, for $d=1,2,3,4$. We just observe that the function is
monotonically decreasing, showing no other remarkable feature than the
small plateau for $d=3,4$. In fact, the values for the three- and
four-dimensional Riemann spheres are very similar. There is a sort of
stabilization of the Casimir force for this number of dimensions.

\bs

\section{The vacuum energy density for cylinders of Riemann surface
section}

As a sort of simplified study of stability of the vacuum energy
density against deformations of the space manifold considered, we
shall now investigate how the preceding values change when we
consider the cylinders, with and without boundary conditions,
which are obtained by adding a flat suplementary dimension to
the Riemann spheres and half-spheres considered above.

\subsection{Dimension d=2}

We start with the case of a two-dimensional cylinder whose sections
are one-dimensional Riemann spheres (no boundaries) or
semicircumferences with Dirichlet and Neumann boundary conditions,
respectively. For the two last cases, we have (for a comprehensible
treatment, see \cite{eenc})
\beq
E_{1,1}^{D,N} =   \frac{\hbar}{4\pi r^2} \, \int_{-
\infty}^{+\infty} dk \, \sum_{n=1,0}^\infty
\sqrt{k^2+\frac{n^2}{r^2}}
=  \frac{\hbar}{4 \sqrt{\pi}r^3} \, \zeta_{1,1}^{D,N} (s=-1/2),
\eeq
being
\beq
 \zeta_{1,1}^{D,N} (s) =   \frac{\Gamma (s-1/2)}{\Gamma (s)} \,
\sum_{n=1,0}^\infty (n^2)^{1/2-s} = \frac{\Gamma (s-1/2)}{\Gamma
(s)} \, \zeta (2s-1).
\eeq
We obtain immediately (no pole appears):
\beq
E_{1,1}^{D,N} =  \frac{\hbar \zeta'(-2)}{4\pi r^3} =-0.0024 \cdot
\frac{\hbar}{r^3}, \  \ \ \ \ E_{1,1} =-0.0048 \cdot
\frac{\hbar}{r^3}.
\eeq

\subsection{Dimension d=3}

For a cylinder made up of ordinary Riemann half-spheres with
Dirichlet or Neumann boundary conditions, what we have is
\beq
E_{2,1}^{D,N} =   \frac{\hbar}{4\pi r^3}  \int_{-\infty}^{+\infty}
dk \, \sum_l^{D,N} \sqrt{k^2+\frac{l(l+1)}{r^2}}
=  \frac{\hbar}{4 \sqrt{\pi}r^4} \, \zeta_{2,1}^{D,N} (s=-1/2),
\eeq
being
\beq
 \zeta_{2,1}^D (s) =   \frac{\Gamma (s-1/2)}{\Gamma (s)} \,
\sum_{l=1}^\infty l [l(l+1)]^{1/2-s}, \ \ \  \zeta_{2,1}^N (s) =
\frac{\Gamma (s-1/2)}{\Gamma (s)} \, \sum_{l=1}^\infty (l+1)
[l(l+1)]^{1/2-s}.
\eeq
We obtain
\bea
 \zeta_{2,1}^D (s) & = &  \frac{\Gamma (s-1/2)}{\Gamma (s)} \left\{
2^{1/2 -s} + \sum_{k=0}^\infty  \frac{\Gamma (3/2-s)}{k! \Gamma
(3/2-s-k)} [\zeta (2s +k -2) -1]\right\}, \nn \\
 \zeta_{2,1}^N (s) & = &  \frac{\Gamma (s-1/2)}{\Gamma (s)} \left\{
2^{3/2 -s} + \sum_{k=0}^\infty  \frac{\Gamma (5/2-s)}{k! \Gamma
(5/2-s-k)} [\zeta (2s +k -2) -1]\right\}.
\eea
Here a pole appears in both cases, with the same residue $-1/(60
\sqrt{\pi})$. Again, one must be here very careful when putting
$s=-1/2$ in these expressions, since one has to expand the zeroes and
the poles
of the gamma and zeta functions in terms of $s+1/2$. After doing so,
and using the following table of values for the derivative of the
zeta function \cite{ejmp2}
\bea
&&\zeta'(-1) = -0.16542115, \ \ \ \ \zeta'(-2) = -0.0304, \ \ \ \
\zeta'(-3) = 0.0054, \nn \\ &&
\zeta'(-4) = 0.0080, \ \ \ \ \zeta'(-5) = 0.00066, \
\ldots
\label{zpr}
\eea
we obtain
\beq
E_{2,1}^D = -0.0042 \cdot \frac{\hbar}{r^4},  \ \ \
E_{2,1}^N
=- 0.0073 \cdot \frac{\hbar}{r^4},  \ \ \ E_{2,1} = -0.0115 \cdot
\frac{\hbar}{r^4}.
\eeq

\subsection{Dimension d=4}

For the case of the four-dimensional cylinder whose sections are
three-dimensional Riemann spheres, we will only consider the
no-boundary situation. Now
\beq
E_{3,1} =   \frac{\hbar}{4\pi r^5}  \int_{-\infty}^{+\infty} dk \,
\sum_{l=1}^\infty (l+1)^2 \sqrt{k^2+l(l+2)}
=  \frac{\hbar}{4 \sqrt{\pi}r^5} \, \zeta_{3,1} (s=-1/2),
\eeq
being
\beq
 \zeta_{3,1} (s) =   \frac{\Gamma (s-1/2)}{\Gamma (s)} \,
\sum_{l=1}^\infty (l+1)^2 [l(l+2)]^{1/2-s} =   \frac{\Gamma (s-
1/2)}{\Gamma (s)} \, \sum_{k=0}^\infty  \frac{(-1)^k \Gamma (3/2-
s)}{k! \Gamma (3/2-s-k)} [\zeta (2s +2k -3) -1] .
\eeq
This case is also finite, no pole appears when computing
$\zeta_{3,1} (-1/2)$. On the other hand, it turns out that the
final value for the energy density is remarkably small:
\beq
 E_{3,1} = -0.0115  \cdot \frac{\hbar}{ r^5},
\eeq
actually, it exactly coincides with the value obtained in the
former case. There seem to be an intriguing stability of the vacuum
energy density for these manifolds at this number of dimensions,
more remarkable because the formulas for the zeta functions leading
to these values look very different indeed (actually, the
three-dimensional case needed a principal part evaluation of the
pole while the four-dimensional case is finite). Such a plateau almost
appeared for Riemann spheres, but here the values obtained are
{\it identical} and the plateau is completely horizontal (see Fig. 2).

\subsection{Dimension d=5}

Finally, we shall  consider the five dimensional cylinder whose
sections are four dimensional Riemann spheres. Here
\beq
E_{4,1} =   \frac{\hbar}{4\pi r^6}  \int_{-\infty}^{+\infty} dk \,
\sum_{l=1}^\infty \frac{(l+1)(l+2) (2l+3)}{6} \sqrt{k^2+l(l+3)}
=  \frac{\hbar}{4 \sqrt{\pi}r^6} \, \zeta_{4,1} (s=-1/2),
\eeq
where
\beq
 \zeta_{4,1} (s) =   \frac{\Gamma (s-1/2)}{3\Gamma (s)} \,
\sum_{l=1}^\infty u^{2(1-s)} (u^2-1/4) [1-9/(4u^2)]^{1/2-s}, \ \ \
\  u=l + 3/2.
\eeq
Using the same methods as before, we obtain,
\beq
 \zeta_{4,1} (s) =
 =   \frac{\Gamma (s-1/2)}{3\Gamma (s)} \, \sum_{k=0}^\infty
\frac{(-1)^k \Gamma (3/2-s)}{k! \Gamma (3/2-s-k)} \left(
\frac{9}{4} \right)^k [\zeta (2s +2k -4,5/2) -\zeta (2s +2k -2,5/2)
/4] .
\eeq
Here we get again the usual pole. The final value is
\beq
 E_{4,1} = -0.0218  \cdot \frac{\hbar}{ r^6}.
\eeq

The results of this section are summarized in Fig. 2, where
the vacuum energy density corresponding to a massless scalar field
living in cylinders whose sections
are Riemann spheres (again manifolds without boundary) is
represented as a function of the space dimension $d$ and in units
of $\hbar \ r^{-(d+1)}$, for $d=2,3,4,5$. A similar monotonic
behavior with dimension as in the case of spheres is observed.
Moreover, we notice an absolute stabilization of the numerical
value of the energy density, which is exactly the same for $d=4$ as for
$d=3$ (sections are two- and three-dimensional Riemann spheres). Also,
the case $d=4$ is somehow special in the sense that it gives a
finite result, no pole arises there. To be remarked as well is the
fact that the Casimir energies for these cylinders are almost
exactly by one order of magnitude smaller (in absolute value) than
the ones obtained for the corresponding Riemann spheres.

\bs

\section{Discussion and conclusions}

We have calculated  in this paper
the vacuum energy density (also called Casimir energy density
\cite{bvw}) corresponding to a massless scalar quantum field living in
different no-boundary universes, as a function of the number of
dimensions. For models of the universe we have chosen mainly
manifolds without boundary, since they seem to be the most accepted
nowadays and, among them,
Riemann spheres, since they are to be counted among the simplest
and most important of these manifolds. However, several manifolds
with a boundary, with Dirichlet and Neumann conditions,
respectively, have also been considered.
 We have  studied the variation of the vacuum energy density with
the dimension of the space, hoping
 to find some characteristic that may singularize some of
the manifolds considered or a particular dimension, among all
of them. For this purpose, we have also compared the results
obtained for spheres with the ones corresponding to cylinders of
spherical section. It is in this respect that an alternative title
of our work could have been: search for the most stable quantum
universe without boundaries (stability in the sense of the vacuum
energy density).

As for the methods of analysis employed, they have consisted in
zeta function regularization  supplemented with
a  simple and natural binomial expansion, which has been shown
to be both rigorous and very well suited for performing the
numerical calculations. The principal part
prescription has been used to deal with the poles that appeared in
several cases.

The most remarkable results of our investigation are the
following. (i) The absence of poles (completely finite result) for
the case of the four-dimensional Riemann sphere and also for the
four-dimensional cylinder of three-dimensional Riemann spherical
section. Also in the low dimensional cases ($d=1,2$) poles are
absent, but this is not true for the intermediate case $d=3$. Of
course also in this case the principal part prescription gives a
well-defined, finite result, but this does not seem to be quite as
satisfactory (for the very purists). (ii) The exact coincidence
 of the results corresponding to a 3d and a 4d cylinder
even if the initial expressions are completely different, and the first
result is obtained only after subtracting the pole. (iii) The fact
that the vacuum energy density for cylinders is by an order of
magnitude smaller than the one corresponding to spheres of the same
dimension. Since all the results are negative, this means that when
the no-boundary manifold is not spherical, the attractive Casimir
force diminishes considerably. (iv) Finally, the elegance,
simplicity and mathematical rigour of our method, as compared with
other approaches, is very remarkable. As discussed before, the
merit for this is to be given  in full to the procedure of zeta
function regularization itself, which is being confirmed as the
most beautiful and useful existing regularization procedure \cite{fin}.

\vspace{5mm}
\noindent{\large \bf Acknowledgments}

I am  grateful to Lars Brink,  Bengt Nilsson, Per Salomonson,
Bo-Sture Skagerstam and Ingemar Bengtsson for discussions and for
the nice atmosphere in G\"oteborg.
This work has been  supported by DGICYT (Spain) and by CIRIT
(Generalitat de Catalunya).

\newpage

\noindent{\large \bf Figure captions}
\bigskip

\noindent{\bf Figure 1}.
The vacuum energy density for Riemann spheres (no-boundary
manifolds) as
a function of the space dimension, $d$, in units of $\hbar \
r^{-(d+1)}$, for $d=1,2,3,4$. The function is monotonically
 decreasing, showing no other remarkable feature than the
small plateau for $d=3,4$. In fact, the values for the three- and
four-dimensional Riemann spheres are very similar. There is a sort of
stabilization of the Casimir force for this number of dimensions.

\bigskip

\noindent{\bf Figure 2}.
The vacuum energy density corresponding to cylinders whose sections
are Riemann spheres (again manifolds without boundary),
as a function of the space dimension $d$ and in units of $\hbar \
r^{-(d+1)}$, for $d=1,2,3,4$. We see the same monotonical behavior
as in the case of the spheres but it is remarkable the clear
stabilization of the numerical value of the energy density which
is exactly the same for $d=4$ as for $d=3$ (cylinders whose
sections are two- and three-dimensional Riemann spheres,
respectively).

\end{document}